\documentclass[a4paper,11pt]{article}

\usepackage{jheppub} 

\usepackage[T1]{fontenc} 
\makeatletter
\def\@fpheader{\relax}
\makeatother

\title{\boldmath  Effective metric in fluid-gravity duality through parallel transport: a proposal}

 \author[a,1]{Shounak De,%
 \note{Present Address: Department of Astronomy and Astrophysics, Tata Institute of Fundamental Research, Homi Bhabha Rd, Mumbai 400005, India}}
 \author[a]{Sumit Dey}
 \author[a]{and Bibhas Ranjan Majhi}
 \affiliation[a]{Department of Physics, Indian Institute of Technology Guwahati, Guwahati 781039, Assam, India}

\emailAdd{shounakde@alumni.iitg.ac.in}
\emailAdd{dey18@iitg.ac.in}
\emailAdd{bibhas.majhi@iitg.ac.in}

\abstract{The incompressible Navier-Stokes (NS) equation is known to govern the hydrodynamic limit of essentially any fluid and its rich non-linear structure has critical implications in both mathematics and physics. The employability of the methods of Riemannian geometry to the study of hydrodynamical flows has been previously explored from a purely mathematical perspective. In this work, we propose a bulk metric in $(p+2)$-dimensions with the construction being such that the induced metric is flat on a timelike $r = r_c$ (constant) slice. We then show that the equations of {\it parallel transport} for an appropriately defined bulk velocity vector field along its own direction on this manifold when projected onto the flat timelike hypersurface requires the satisfaction of the incompressible NS equation in $(p+1)$-dimensions. Additionally, the incompressibility condition of the fluid arises from a vanishing expansion parameter $\theta$, which is known to govern the convergence (or divergence) of a congruence of arbitrary timelike curves on a given manifold. In this approach Einstein's equations do not play any role and this can be regarded as an {\it off-shell} description of fluid-gravity correspondence. We argue that {\it our metric effectively encapsulates the information of forcing terms in the governing equations as if a free fluid is parallel transported on this curved background}. We finally discuss the implications of this interesting observation and its potentiality in helping us to understand hydrodynamical flows in a probable new setting.}

\begin{document} 
\maketitle
\flushbottom

\section{Introduction}
\label{sec:intro}
The non-relativistic fluid flow is dictated by the well-known incompressible Navier-Stokes (NS) equations
\begin{align}
\dot{\vec{v}} + \vec{v} . {\nabla} \vec{v} \, + & \, \vec{\nabla} P - \eta \nabla^2 \vec{v} = 0~;
\nonumber \\
\vec{\nabla} . \vec{v} &= 0~,
\label{0.01}
\end{align}
where $\vec{v}$ is the fluid velocity field, $P$ the fluid pressure and $\eta$ the kinematic viscosity. The NS equation of fluid dynamics has played a crucial role in both mathematics and physics. The NS differential equation which exhibits a rich non-linear structure has been seen to describe a wide variety of natural phenomena \cite{Landau}. The subject of hydrodynamics has been studied extensively for centuries now, yet many open ended questions remain to be answered. For instance, a consistent mathematical theory of the phenomenon of turbulence and the existence problems for the smooth solutions of hydrodynamic equations (\ref{0.01}) of a three-dimensional fluid are still wide open. 

A simpler yet a rigorous mathematical model for fluid dynamics is the hydrodynamics of an ideal fluid, i.e., an Euler fluid (described by (\ref{0.01}) with $\eta = 0$). According to Arnold \cite{Arnold}, from the mathematical point of view, ``a theory of such a fluid filling a certain domain is nothing but a study of geodesics on the group of diffeomorphisms of the domain that preserve volume elements. The geodesics on this (infinite-dimensional) group are considered with respect to the right-invariant Riemannian metric.'' From this purely mathematical perspective, the Euler equations of fluid dynamics on a compact \textit{n}-dimensional manifold $\mathcal{M}$ can be regarded as the equation of geodesics on the Lie group \textit{S Diff ($\mathcal{M}$)} of all diffeomorphisms on the manifold $\mathcal{M}$ preserving the volume form $V$. The description of ideal hydrodynamical flows by means of geodesics of the right-invariant metric allows one to employ the methods of Riemannian geometry to the study of flows. 

In our paper, we try to understand to what degree the NS equation (\ref{0.01}) can have such a manifold interpretation, partially motivated by investigations and advances in the domain of the fluid-gravity correspondence. One of the earliest works regarding this interesting connection between the equations of gravity and hydrodynamics appeared in the doctoral thesis of Damour \cite{Damour:1979wya}, wherein there are suggestions of a relation between horizon and fluid dynamics. This work contains an expression now known as the Damour-Navier-Stokes (DNS) equation and is known to govern the geometric data on any null surface. The same equation is also obtained in terms of coordinates adapted to a null surface \cite{Padmanabhan:2010rp} by projecting the Einstein's equations of motion. Moreover, a corresponding action formulation of the same has been greatly detailed in \cite{Kolekar:2011gw}. A connection in this regard has also been obtained in the membrane paradigm approach by Price and Thorne in \cite{Price:1986yy}. In the AdS/CFT context, it has been shown that the dissipative behaviour of an AdS black hole agrees with the hydrodynamics of the holographically dual CFT. This has been studied extensively and important works in this regard include \cite{Bhattacharyya:2008kq, Bhattacharyya:2008jc, Bhattacharyya:2008ji, Policastro:2002se}. More recently in a cut-off surface approach by Bredberg et al. \cite{Bredberg:2011jq}, it has been shown by explicit construction that for every solution of the incompressible NS equation in $(p+1)$-dimensions, there is a uniquely associated dual solution of the vacuum Einstein equations in $(p+2)$-dimensions. This cut-off surface approach has been applied in various cases, see \cite{Huang:2011he, De:2018zxo, Zhang:2012uy}. For example, it was extended for higher curvature gravity theories \cite{Chirco:2011ex, Bai:2012ci, Cai:2012mg, Zou:2013ix, Hu:2013lua} as well as for the AdS \cite{Cai:2011xv, Huang:2011kj} and dS \cite{Anninos:2011zn} gravity theories (for other theories, like black branes, see \cite{Ling:2013kua}). Very recently, two of the authors of this paper showed in \cite{De:2018zxo} that an incompressible DNS-like equation can be obtained in the cut-off surface approach. In this case the obtained metric is a solution of Einstein's equations of motion in the presence of a particular type of matter. Also a corresponding relativistic situation has been discussed extensively in \cite{Eling:2012ni}. Symmetries of the vacuum Einstein equations have been exploited to develop a formalism for solution generating transformations of the corresponding NS fluid duals in \cite{Berkeley:2012kz}. The fluid description on the Kerr horizon has also been explored extensively in \cite{Lysov:2017cmc} (see \cite{Wu:2013aov} for the isolated horizon case). For extensive reviews of the fluid-gravity correspondence, refer to \cite{Padmanabhan:2009vy, Hubeny:2010wp, Rangamani:2009xk}. 

In this paper, we present an interesting observation which could help to understand if the fluid-gravity correspondence can serve as a window to view hydrodynamical flows in a new setting. Here we propose a metric which is constructed with the help of the scaling symmetry of the incompressible (D)NS equation{\footnote{For the details of the scaling symmetry of (D)NS, refer to appendix A of \cite{De:2018zxo}.}}. We show that the parallel transport equation of the fluid velocity along its own direction on a timelike hypersurface in a manifold specified by this metric yields (D)NS. The incompressibility condition naturally arises from the vanishing of the expansion parameter  corresponding to the fluid velocity vector. It must be noted that our whole approach, as well as the analysis that follows, is completely different from existing works in this direction. Moreover, here we do not need to consider the Einstein's equations of motion explicitly. Therefore this new approach can be regarded as an {\it off-shell} description.

Our basic organization of the paper is roughly as follows. In section \ref{para}, we begin by first writing down a bulk metric in $(p+2)$-dimensions on which we consider the parallel transport of an appropriately defined velocity vector field. We then show that the projection of the parallel transport equations onto a timelike induced hypersurface require that the incompressible fluid dynamical (D)NS equations be satisfied in $(p+1)$-dimensions. In section \ref{exp}, we show that the incompressibility condition of the fluid derives from a vanishing expansion parameter $\theta$ when projected onto the same timelike induced hypersurface. The next section is dedicated to show that our proposed metric is inherently curved and a solution of the vacuum Einstein equations of motion. In section \ref{disc}, we finally discuss the consequences of this interesting observation and how it lends a different perspective in viewing the (D)NS fluid dynamical equation. The order-wise calculations of the projected parallel transport equations and that of the expansion parameter $\theta$ is explicitly shown in the appendices \ref{App1} and \ref{App3}.

The \textit{notations} used throughout the paper are clarified as follows: all lowercase Latin letters denote the bulk spacetime co-ordinate indices and run from $a, b = 0, \dots, p+1$. The uppercase Latin letters denote the transverse co-ordinates (i.e., the angular sector of the metric) and they run from $A, B = 1, \dots, p$. The Greek letters denote the co-ordinates on the flat timelike induced $r = r_c$ slice and they run from $\mu, \nu = 0, \dots, p$. 


\section{\label{para}Parallel transport to fluid dynamics} 
In this section, we construct a metric which effectively captures the features of a viscous fluid such that this viscous fluid is equivalently represented by a free fluid parallel transported along its own flow direction on a specific hypersurface in this background. In other words, we want to find an equivalent gravity description such that the metric coefficients encapsulates the information of the forcing terms (the ones arising due to pressure gradient and viscous dissipation) which are acting on a fluid described on a flat space and reinterpreted as if a free fluid is flowing on a curved background. Here we mainly concentrate on a non-relativistic, viscous fluid whose governing equations is the NS equation (\ref{0.01}) accompanied by the incompressibility condition.

\subsection{The metric}
To construct the metric, we shall take help of the well known scaling symmetry of the incompressible NS equation (\ref{0.01}) which we shall briefly state below. Now, if the amplitudes of its solution space $(v_A, P)$ is scaled down by the parameter $\epsilon$:
 \begin{align}
 v_A^\epsilon(x^A,\tau)&=\epsilon v_A(\epsilon x^A,\epsilon^2 \tau)~;
 \nonumber 
 \\
 P^\epsilon(x^A,\tau)&=\epsilon^{2}P(\epsilon x^A,\epsilon^2 \tau)~,
 \label{1.00}
 \end{align}
then the NS equation remains invariant under the above scaling transformation, thus generating a family of solutions parameterized by $\epsilon$ from the original solution space. The hydrodynamic scaling of the spatial and time derivatives along with the pair $(v_A,P)$ follows as:
 \begin{align}
 v_A \sim \mathcal{O} (\epsilon)\,, \quad  P \sim \mathcal{O} (\epsilon^2)\,, \quad \partial_A  \sim \mathcal{O} (\epsilon)\,,\quad \partial_{\tau} \sim \mathcal{O} (\epsilon^2)~.
 \label{1.001}
 \end{align}
A review of the same can be found in \cite{Bredberg:2011jq} (also see appendix A of \cite{De:2018zxo} for a detailed derivation of the scale invariance of the incompressible DNS equation). Like the method adopted originally in \cite{Bredberg:2011jq}, the hydrodynamic scaling parameter $\epsilon$ is taken to be the perturbative parameter in which the metric is expanded and will be a recurrent feature of our analysis. Here we propose the following metric:
\begin{align}
ds_{p+2}^2 &= g_{a b}dx^adx^b = -rd\tau^2 + 2d\tau dr + dx_Adx^A \nonumber\\
&+ \bigg{(}\frac{2 a_1}{r_c} \partial_A P + 2 a_2 \partial^2 v_A + \frac{2 a_3}{r_c} \partial_A v^2 \bigg{)} dx^A dr + \mathcal{O}(\epsilon^4)~.
\label{1.01} 
\end{align}
The metric is constructed in such a way that the leading order base metric at $\mathcal{O}(\epsilon^0)$ is in flat Rindler form in (ingoing) Eddington-Finkelstein coordinates and serves as the background metric $g_{a b}^{(0)}$. The next metric coefficients appear at $\mathcal{O}(\epsilon^3)$ which acts as a perturbation in the third order in $\epsilon$ and we denote it by $h_{a b}^{(3)}$. The metric construction is such that the induced metric on a timelike induced slice $r=r_c$ is flat, i.e.
\begin{align}
\gamma_{\mu \nu} dx^{\mu} dx^{\nu} = - r_c d \tau^2 + dx_A dx^A \,.
\label{1.01a}
\end{align}
We also note that the velocity $v_A(x^A, \tau)$ and pressure $P(x^A, \tau)$ fields are independent of the radial coordinate $r$. We shall observe that the above metric (\ref{1.01}) correctly incorporates the information of the forcing terms in the NS equation on the $r=r_c$ hypersurface through the metric coefficient at $\mathcal{O}(\epsilon^3)$. Therefore the present proposed curved spacetime, on the timelike hypersurface, acts as the gravity dual of the governing non-relativistic fluid equations on flat space. This we shall show below by using the parallel transport equations of a free fluid along its flow direction, projected onto the timelike induced $r = r_c$ hypersurface.

\subsection{\label{paraorder}Parallel transport and fluid equation}
Since the calculation will be done by projecting everything onto the $r=r_c$ timelike hypersurface, we first evaluate the projectors, defined by 
\begin{equation}
h_{ab} =g_{ab}-n_an_b~,
\label{3}
\end{equation}
where $n_a$ is the unit normal to the $r=r_c$ surface, satisfying $n_an^a=+1$ (spacelike).
The projectors corresponding to the leading order background metric are listed as under:
\begin{align}
&h_{\tau \tau} = -r\,, \quad h_{\tau r} = 1\,, \quad h_{\tau A} = h_{r A} = 0\,,\quad h_{r r} = -1, \quad h_{A B} = \delta_{A B}\,.
\label{1.02} 
\end{align}
The corresponding contravariant components of the above projectors are listed as follows:
\begin{align}
&h^{\tau \tau} = -\frac{1}{r}\,, \quad h^{\tau r} = h^{\tau A} = 0 \,, \quad h^{r r} = h^{r A} = 0\,, \quad h^{A B} = \delta^{A B}\,.
\label{1.02a}
\end{align}

Now, the notion of ``straight'' paths in a curved space are those curves $x^a(\tau)$ for which the tangent vector $v^a = (dx^a/d\tau)$ should be parallel transported along the same curve. Here we consider that there is no fluid flow along the radial direction. In this case the bulk velocity vector field components are $v^a = (1,0,v^A)$. Now, parallel transport requires that the directional derivative of $v^b$ along $v^a$ (i.e. along the its own flow direction) must vanish, i.e. we must have the satisfaction of the following relation:
\begin{align}
v^a \nabla_a v^b = 0 \,.
\label{1.03}
\end{align}
The projection of (\ref{1.03}) onto the timelike induced hypersurface $r = r_c$ (since we are interested on this surface) yields the relation:
\begin{align}
h_{b c} v^a \nabla_a v^b \bigg|_{r = r_c} = 0 \,.
\label{1.04}
\end{align}
We shall show that the satisfaction of the above equation identically upto $\mathcal{O}(\epsilon^3)$, i.e. upto the order the metric (\ref{1.01}) is presented, generates the (D)NS fluid dynamical equations.

Let us now start expanding Eq. (\ref{1.04}) for different indices. First setting the free index $c = \tau$ in (\ref{1.04}), we obtain:
\begin{align}
h_{b \tau} v^a \nabla_a v^b \bigg|_{r = r_c} =& - r \big{(}\Gamma_{\tau \tau}^{\tau} + 2 \Gamma^{\tau}_{\tau A} v^A + \Gamma^{\tau}_{A B} v^A v^B\big{)} + \Gamma^{r}_{\tau \tau} + 2 \Gamma^{r}_{\tau A} v^A + \Gamma^{r}_{A B} v^A v^B \,.
\label{1.05a} 
\end{align}
We proceed to check whether the right hand side (RHS) of (\ref{1.05a}) vanishes identically order by order in the hydrodynamic expansion parameter $\epsilon$. It is found that (\ref{1.05a}) vanishes identically upto $\mathcal{O}(\epsilon^3)$ and (\ref{1.04}) is satisfied trivially for $c = \tau$ (See appendix \ref{App1} for a detailed order by order calculation). 
Next, on setting the free index $c = A$ in (\ref{1.04}), we obtain:
\begin{align}
h_{b A} v^a \nabla_a v^b \bigg|_{r = r_c} &= \delta_{A B} \big{(} \partial_{\tau} v^B + v^C \partial_C v^B + \Gamma^{B}_{\tau \tau} + 2 \Gamma^{B}_{C \tau} v^C + \Gamma^{B}_{C D} v^C v^D \big{)} \,.
\label{1.06a} 
\end{align}
As before, we again proceed to check the RHS of (\ref{1.06a}) order by order in the hydrodynamic expansion parameter $\epsilon$. Here we find that upto order $\epsilon^2$, it is automatically satisfied, whereas the above expression, at order $\epsilon^3$, yields
\begin{align}
\mathcal{O}(\epsilon^3):~ &h_{b A} v^a \nabla_a v^b \bigg|_{r = r_c}= \partial_{\tau} v_A + v^C \partial_C v_A + \frac{a_1}{2} \partial_A P + \frac{a_2}{2} r_c \partial^2 v_A + \frac{a_3}{2} \partial_A v^2 \,. 
\end{align}
Again, refer to appendix \ref{App1} for a detailed order-wise calculation. 
So the satisfaction of Eq. (\ref{1.04}) requires that the above must vanish at this order, i.e. 
\begin{equation}
\partial_{\tau} v_A + v^C \partial_C v_A + \frac{a_1}{2} \partial_A P + \frac{a_2}{2} r_c \partial^2 v_A + \frac{a_3}{2} \partial_A v^2=0 \,.
\label{1.06b}
\end{equation}
Now with the identification of the constants $a_i$'s in the metric (\ref{1.01}) to be:
\begin{align}
a_1 = 2 \,, \quad a_2 = -2 \,, \quad a_3 = 0 \,,
\end{align}
and identifying the (kinematic) viscosity as $\eta = r_c$, Eq. (\ref{1.06b}) yields:
\begin{align}
\partial_{\tau} v_A + v^C \partial_C v_A + \partial_A P - \eta \partial^2 v_A  =0 \,. 
\label{1.07}
\end{align}
Note that the above one is the incompressible NS equation. In the next section we shall discuss how one can obtain the incompressibility condition in the gravity dual paradigm in the present setup.

Next if we choose $a_1=2$, $a_2=-2$ and $a_3=1$, then Eq. (\ref{1.06b}) reduces to the following form:
\begin{align}
\partial_{\tau} v_A + v^C \partial_C v_A + \frac{1}{2} \partial_A v^2 + \partial_A P - \eta \partial^2 v_A = 0 \,,
\label{1.08}
\end{align}
which is seen to be the incompressible Damour-Navier-Stokes (DNS) equation, a gravity dual of which has been extensively studied recently in \cite{De:2018zxo}. Thus, it is seen that for the metric (\ref{1.01}), the parallel transport equation when projected onto a timelike induced flat hypersurface (\ref{1.04}), leads to (D)NS fluid dynamics on this surface. Although as seen from (\ref{1.08}) again, the incompressibility condition is yet to be obtained, which we shall tackle in the next section.

\section{\label{exp}Expansion parameter and incompressibility}
The Raychaudhuri equation governs the convergence (or divergence) of a congruence of arbitrary timelike curves. This is essentially done by determining the expansion parameter $\theta$ of the congruence. With the already defined vector field $v^a = (1,0,v^A)$, we define the expansion parameter as
\begin{align}
\theta = h^{a b}\nabla_a v_b = h^{a b} (\partial_a v_b - \Gamma^{c}_{a b} v_c) = h^{a b} (\partial_a v_b + r \Gamma^{\tau}_{a b} - \Gamma^{r}_{a b} - \Gamma^{A}_{a b} v_A) \,,
\label{1.09a}
\end{align}
where the projectors $h_{ab}$ are defined as given by (\ref{3}). Now for incompressible flows, we must have a vanishing $\theta$. As before, we proceed to check the right hand side of (\ref{1.09a}) in an order-wise manner in the hydrodynamic expansion parameter $\epsilon$. An explicit calculation shows that except at $\mathcal{O}(\epsilon^2)$, all other orders upto $\epsilon^3$ vanish identically. The $\mathcal{O}(\epsilon^2)$ term in (\ref{1.09a}) is simply $\sim\partial_Av^A$ (see appendix \ref{App3} for an explicit order-wise evaluation of (\ref{1.09a})). Therefore, to satisfy the vanishing of $\theta$ upto order $\epsilon^3$, we need to impose the following condition:
\begin{align}
\partial_A v^A = 0 \,, 
\label{1.10}
\end{align}
which is the required incompressibility condition. In the context of fluid dynamics, we speak of incompressible flows when there is no noticeable compression or expansion of the fluid (see \cite{Landau}). Therefore, it seems very natural that the incompressibility condition emerges from a vanishing expansion parameter. This is what we have shown in this section.

\section{More about proposed metric}
The proposed metric (\ref{1.01}) is seen to effectively capture the features of an incompressible, viscous fluid such that this (D)NS fluid is equivalently represented by a free fluid parallel transported along its own flow direction on a specific hypersurface in this background, as explicitly detailed in the preceding sections \ref{para} and \ref{exp}. This metric indeed represents a truly curved manifold as established by the fact that the Riemann curvature tensor is seen to have non-vanishing contributions at $\mathcal{O}(\epsilon^3)$ itself, for example:
\begin{align}
R_{A \tau r \tau}^{(3)} = \frac{1}{2} \bigg( \frac{a_1}{r_c} \partial_A P + a_2 \partial^2 v_A + \frac{a_3}{r_c} \partial_A v^2 \bigg) \,.
\label{4.02}
\end{align}
Now that the metric is established to be representative of a true curved spacetime, we need to check whether its dynamics is governed by the Einstein field equations. Crucially, the metric (\ref{1.01}) is a solution to the vacuum Einstein field equations upto $\mathcal{O}(\epsilon^3)$, i.e., upto the order in which the bulk metric is presented. In other words, the Ricci tensor components vanish identically upto $\mathcal{O}(\epsilon^3)$, with non-vanishing contributions arising only at $\mathcal{O}(\epsilon^4)$ and higher. That is all components of the Einstein tensor obey
\begin{align}
G_{r r}, \, G_{r \mu}, \, G_{\mu \nu} \sim \mathcal{O}(\epsilon^4) \,,
\label{4.01}
\end{align}
and are nonsingular for finite values of $r$. 

It is well known that a suitable coordinate transformation can always be chosen such that a given metric has the Cartesian form at a given event (say $\mathcal{P}$) and its first derivatives vanish at that event. Such a construction of a coordinate system around an event $\mathcal{P}$ is known to be called a \textit{local inertial frame} at $\mathcal{P}$. The fact that the second derivatives of the metric survive after making the coordinate transformation to a local inertial event at $\mathcal{P}$ can be used to understand features arising from the curvature. This is explicitly seen by virtue of a coordinate transformation, sometimes called the Riemann normal coordinates, which gives the metric in these coordinates to be:
\begin{align}
g_{a b} = \eta_{a b} - \frac{1}{3} (R_{a c b d} + R_{b c a d}) x^c x^d~.
\label{4.02}
\end{align}
This result explicitly shows that a coordinate system can be set up at some event $\mathcal{P}$ such that the metric has the Cartesian form with the second derivatives of the metric traded off for the components of the Riemann curvature tensor. For a detailed derivation of the same, refer to the discussion in Section 5.3.1 of \cite{Paddybook}. 

In a similar vein, we present a construction of a coordinate system about a given event $\mathcal{P}$ (which is taken to be a region about $r = 0$ where $\Gamma_{b c}^{a \, (3)}$'s vanish) such that the deviation of the metric (\ref{1.01}) from the Rindler (in our case) is given by the curvature tensor. To this end, we see that the bulk metric derives from the Rindler background ($g_{a b}^{(0)}$) via a transformation of the coordinates of the following form:
\begin{align}
\label{4.03}
\tilde{x}^A &= x^A + \lambda \, \partial_r \Gamma^{A \, (3)}_{\tau \tau} \, r \, , 
\nonumber
\\ 
\tilde{\tau} &= \tau \, , 
\nonumber
\\
\tilde{r} &= r \, ,  
\end{align} 
such that the event $\mathcal{P}$ is taken to be the origin of both the coordinate systems and where $\lambda$ is simply a constant that shall be fixed in due course. Using (\ref{A1.2}) and (\ref{4.03}), it is easy to see that:
\begin{align}
d\tilde{x}^A = dx^A + \frac{\lambda}{2} \bigg[\bigg( \frac{a_1}{r_c} \partial^A P + a_2 \partial^2 v^A + \frac{a_3}{r_c} \partial^A v^2 \bigg)\bigg] dr + \mathcal{O}(\geq \epsilon^4) \,. 
\label{4.04}
\end{align}
Imposing the above coordinate transformation onto the flat background Rindler metric, we obtain:
\begin{align}
d\tilde{s}_{p+2}^2 &= -\tilde{r}d\tilde{\tau}^2 + 2d\tilde{\tau} d\tilde{r} + d\tilde{x}_Ad\tilde{x}^A \nonumber\\
&= -rd\tau^2 + 2d\tau dr + dx_Adx^A \nonumber\\
&+ \lambda \bigg{(}\frac{a_1}{r_c} \partial_A P + a_2 \partial^2 v_A + \frac{a_3}{r_c} \partial_A v^2 \bigg{)} dx^A dr + \mathcal{O}(\epsilon^4)~.
\label{4.05}
\end{align}
On setting $\lambda = 2$, we identically obtain our proposed bulk metric (\ref{1.01}) upto the required order, i.e., $\mathcal{O}(\epsilon^3)$. Thus our bulk metric construction (which is a true curved manifold) can also be seen to be an expression of a curved background expressed in a {\it locally Rindler frame}. 
 
\section{\label{disc}Discussions and Comments}
To summarise our calculations, we first construct a bulk metric in $(p+2)$-dimensions in which the background $g_{a b}^{(0)}$ is flat Rindler space onto which a perturbation $h_{a b}^{(3)}$, parametererized by the velocity and pressure fields $v_A(x^A, \tau)$ and $P(x^A, \tau)$ of an incompressible fluid, kicks in at the third order in $\epsilon$. Now, on defining a bulk velocity vector field of the form $v^a = (1, 0, v^A)$, we consider the parallel transport of $v^a$ along the integral curves to this vector field on the manifold defined by the spacetime metric (\ref{1.01}) as given by (\ref{1.03}). Next, we consider the projection of these equations of parallel transport onto the flat timelike induced $r = r_c$ slice as shown by (\ref{1.04}). Crucially, the satisfaction of these projected parallel transport equations identically upto $\mathcal{O}(\epsilon^3)$ (i.e., upto the order the bulk metric is presented) require that the incompressible fluid dynamical (D)NS equations be satisfied at the same order. Interestingly, the incompressibility condition arises from an identically vanishing expansion parameter projected onto the same flat $r = r_c$ slice. As far as we know, this way of interpreting the (D)NS fluid equations, in the context of fluid-gravity correspondence, has not been done earlier.

This observation could lend an interesting perspective in our viewpoint of the incompressible Navier-Stokes fluid dynamical equations. Rewriting the incompressible NS equation in the $\textbf{F} = m \textbf{a}$ form:
\begin{align}
\partial_{\tau} v_A + v^B \partial_B v_A = - \partial_A P - \eta \partial^2 v_A \,,
\label{2.2}
\end{align}
where the left hand side (LHS) is the usual fluid convective derivative and the RHS contains the forcing terms arising out of pressure gradient and dissipation due to viscosity. On observing the relation obtained in (\ref{1.06a}), it is seen that these forcing terms essentially arise from the evaluation of the Christoffel symbols defined for the metric (\ref{1.01}) and is a direct consequence of the perturbation $h_{a b}^{(3)}$. Thus, if we were to ``turn off'' this perturbation (i.e., essentially setting $a_1 = a_2 =  a_3 = 0$ in (\ref{1.01})), we would essentially obtain a forcing-free fluid. So the picture here is that a viscous, incompressible fluid residing in flat space-time is essentially equivalent to a free fluid residing in a curved space-time manifold defined by a unique choice of the metric (\ref{1.01}). This seems to be in parallel to the interpretation that an interacting particle with gravitational field in a flat space-time can be equivalently pictured to be a free particle residing in an appropriate curved space-time manifold, which is broadly discussed in \cite{Paddybook} (see the discussion in Section $3.3$ of this book). 

A more ambitious interpretation can be the following. We have noted that our proposed metric correctly accounts for the forcing terms of the (D)NS equation. Therefore the present metric representation can be interpreted as an {\it equivalent theory of the viscous fluid motion}. This is similar to the equivalence principle of gravity where an accelerated frame locally mimics gravity. Therefore, to better understand the (D)NS equation and the properties of a viscous fluid, this metric could probably play a major role. Moreover, any calculation (for a free fluid) on this background reflects the effects of forcing terms and one will be able to extract features of a viscous fluid, which may be very hard to obtain directly from the (D)NS itself as it is a highly non-linear equation. Therefore the present analysis can be a complete {\it geometrical description} of the (D)NS equation. We take this as a suggestive interpretation, rather than a conclusive one. In addition to that, it may be noted that in our present analysis Einstein's equations do not play any role which were the main input in all earlier interpretations \cite{Bhattacharyya:2008jc}--\cite{Wu:2013aov} of (D)NS as dual to a gravitational theory. Hence we designate this approach as an {\it off-shell} fluid-gravity duality scenario. 

There are certain important issues of this approach that can be mentioned here. Firstly, regarding the question of uniqueness of the metric (\ref{1.01}) and how the inclusion of other metric coefficients could affect our analysis. In this regard we have to note that the metric (\ref{1.01}) is exactly what is needed for the (D)NS fluid data to be a condition of parallel transport on this curved background. It has been checked that the inclusion of metric coefficients of the form $d\tau dx^A$ do not affect our analysis (also, there are other possible terms which do not affect the present interpretation of (D)NS as a parallel transport equation). Hence, in principle, such terms could be added to the metric and consequently the metric (\ref{1.01}), strictly in this sense, would then be non-unique. 
But it is to be noted that we are interested in the (D)NS equation which is an unique fluid equation in the hydrodynamic limit (as other higher order terms in the equation are traced out at this limit). So the entire dynamics of the fluid is reflected by this equation and hence a metric description of such a equation is completely determined by the minimum required terms in the metric. Any other correctional terms, allowed by the scaling argument in the metric, are regarded as ``redundant terms'' as far as the dynamics is concerned.  In our prescription, such redundant terms do not affect the dynamics and the physical results we intend to achieve. The idea is similar to the arbitrariness up to the total derivative or constant term in the construction of a Lagrangian for a system. We know that such a term does not affect the equations of motion (i.e. the dynamics) of the system and hence one can just neglect those terms. Thus, at the very outset, the metric is kept devoid of such redundant terms and we work with the simplest metric that would lead to the desired physical results following our prescription. In this sense, the question of uniqueness of the constructed metric (\ref{1.01}) does not have strong physical relevance. This feature of the metric can be related to the gauge symmetry of the electromagnetic fields. In this case, we know that the Maxwell's equations remain invariant under the change of the vector potential field up to some additive term which may be a constant or the derivative of a scalar function. So if we are only concentrating on the (D)NS equation, then our metric (\ref{1.01}) is defined up to some extra allowed metric coefficients which do not change this fluid equation.  

Another question that could possibly arise is: why solely the (D)NS equations have been singled out as a condition of parallel transport on a curved geometry. In principle one could have correctional term(s) to the (D)NS fluid equations and analogously recover corresponding metric(s). The resolution of this argument is pretty simple. We are operating in the regime of the hydrodynamic limit ($\epsilon \rightarrow 0$) and the non-relativistic incompressible (D)NS equations are the precise and universal outcome of such a limit \cite{De:2018zxo}. In this limit, all reasonable types of corrections to the forcing part are scaled away and the incompressible NS equations universally govern the hydrodynamic limit of essentially any fluid. The possibility of changing the right-hand side of the NS equation (\ref{2.2}) at will is nullified in this limit. It is this hydrodynamic limit of a fluid that we have matched to a metric formulation and consequently as a condition of parallel transport on this curved space.

Also, we have stressed on numerous occasions in our paper that the Einstein field equations are not involved in our construction. This is the fact that sets our work on a different footing from the existing holographic approach to the fluid-gravity correspondence where explicit use of Einstein's equations of motion are made. In this regard, let us point out that there is no hard and fast rule that the Einstein's equations of motion has to be the only guiding principle to handle the NS equation in a metric representative construction. The interpretation of Einstein's equations as NS equation on a time-like surface, which has been the main idea adopted till now in various cases, can be regarded as one of the ways to fluid-gravity duality. But there can be other ways to encounter such an issue. This is precisely addressed here and we have found another way through the equations of parallel transport as our guide, instead of the Einstein's field equations.  

Therefore in a nut-shell, once we have established that the NS data governs the hydrodynamic limit of essentially any fluid, we use the hydrodynamic scaling parameter $\epsilon$ as the expansion parameter to write down the metric (\ref{1.01}). The objection about the uniqueness of the metric does not hinder our observation in any way -- corresponding to the incompressible NS equations in the hydrodynamic limit, the metric (\ref{1.01}) is exactly what is needed for the fluid data to be a condition of parallel transport on this curved background. The metric is shown to be a solution to the vacuum Einstein field equations. The field equations, in our case, do not guide the construction of the geometry as in the holographic construction. Our work significantly deviates from the existing approaches to the connection between gravity and hydrodynamics and could potentially shed very interesting light in our understanding of hydrodynamical flows as we have discussed in the paper. 

An important question worth asking is whether the above calculations hint at a manifold interpretation of the incompressible (D)NS fluid equations. In other words, can the (D)NS equations of fluid dynamics be regarded as the equation of geodesics on the Lie group \textit{S Diff}$(\mathcal{M})$ of all diffeomorphisms on the manifold $\mathcal{M}$ defined by the metric (\ref{1.01}). There may an important future aspect of our result. Since we interpreted the (D)NS equation as free parallel transport equation (geodesics) of a fluid on a curved background, probably it can be a path to obtain the action representation of (D)NS equation. The idea is similar to the writing of an action of a free particle on a curved background which, in turn, is identical to a particle interacting with gravity. 

Another important discussion is the systematic generalization of the bulk metric (\ref{1.01}) to higher orders in the hydrodynamic expansion parameter $\epsilon$, in spirit of the reconstruction executed in \cite{Compere:2011dx} for the vacuum solution presented in \cite{Bredberg:2011jq} to arbitrary order. In the work by \cite{Compere:2011dx}, the extension of the vacuum solution in \cite{Bredberg:2011jq} to arbitrary order is carried out through the satisfaction of suitable integrability conditions. The validity of the integrability conditions is ensured by invoking the Bianchi identity and the Gauss-Codazzi equations at the required order. From the perspective of the dual fluid, this constraint reduces to the incompressibility condition at $\mathcal{O}(\epsilon^2)$ and to the exact NS equation at $\mathcal{O}(\epsilon^3)$. The higher order corrections to the bulk metric solution subsequently lead to corrections to the incompressibility condition at relevant even orders in $\epsilon$, while at higher odd orders it amounts to adding corrections to the NS equation. A similar reconstruction of the metric (\ref{1.01}) to arbitrary order could be executed under the purview of the formalism presented in our paper. As we have already seen, the forcing terms in the exact NS equation essentially arise from the evaluation of the Christoffel symbols defined for the metric (\ref{1.01}) and is a direct consequence of the metric coefficient $h_{rA}^{(3)}$. In principle, one could have higher order corrections to the metric solution (\ref{1.01}), say at order $\epsilon^n$, encoded in the corresponding metric co-efficient $h_{rA}^{(n)}$. Consequently such correction terms in the metric, in the parallel transport formalism, will lead to corrections to the incompressibility condition at higher even orders in $\epsilon$ ($\epsilon^4, \epsilon^6, \dots $), while at higher odd orders ($\epsilon^5, \epsilon^7, \dots $) it will amount to adding corrections to the NS equation of fluid dynamics. The allowed metric corrections will be constrained by invoking manifold properties, possibly the Bianchi identity and constraints on the expansion parameter $\theta$. On such grounds, the metric will still continue to be a solution of the vacuum Einstein field equations and undergo consequent generalization to higher orders. 

In addition to the above discussions, there are several pertinent questions regarding possible connections between the formalism presented in this paper with the existing holographic approaches to fluid-gravity duality. For example, it is to be checked whether the proposed metric (\ref{1.01}) bears any direct relationship to metrics in the various fluid-gravity scenarios established over the past few decades. In the cut-off surface approach to fluid-gravity duality as developed in \cite{Bredberg:2011jq}, the metric is seen to be derived by the action of diffeomorphisms on a flat Rindler spacetime as shown in \cite{Compere:2011dx}. The transformations include a constant boost followed by a constant linear shift of the radial co-ordinate $r$ and an associated re-scaling of the time co-ordinate $\tau$, applied in either order on Rindler spacetime. Promoting the velocity $v^A$ and pressure $P$ fields to depend arbitrarily on the co-ordinates $x^A$ and treating them as small fluctuations around the background in the hydrodynamic limit yields the required metric as given in \cite{Bredberg:2011jq}. It would be really interesting to see whether our metric (\ref{1.01}) could be derived by the action of diffeomorphisms on a given background metric. Moreover, the existing fluid-gravity approaches need to be explored for a potential interpretation along the lines of the manifold properties as discussed in this paper.

Finally, it would also be interesting to explore the consequences of repeating our calculations around spacetimes with a different sort of horizon (instead of Rindler one), like pure de-Sitter cosmological horizon. In this case one can consider a timelike hypersurface near to the cosmological horizon. This has been attempted in \cite{Anninos:2011zn} through the earlier existing approaches. Moreover, for just mathematical interest the same has been done for the spacelike slices foliating the region outside the future horizon of the static patch. In both the cases, the NS equation was recovered, in the hydrodynamic limit. Interestingly, for the timelike case the viscosity (kinematic) coefficient is positive while for the spacelike case it is negative. What would happen if we do everything by our present approach by taking the seed metric as these surfaces would be very interesting to investigate. Similarly, the extension to this present parallel transport approach in the context of Petrov-type constructions \cite{Huang:2011he} would also be worth to look at. We leave such discussions open and up for careful consideration in the near future.

\acknowledgments
We are immensely grateful to Shiraz Minwalla and Yakov Landau for illuminating conversations. The anonymous referee is also greatly acknowledged for posing several constructive questions and comments. 

\vskip 5mm
\section*{Appendices}
\appendix
\section{{\label{App1}}Evaluation of the projected parallel transport equations (\ref{1.05a}) -- (\ref{1.06a}) upto $\epsilon^3$ order}
In order to explicitly evaluate Eq. (\ref{1.05a}) and Eq. (\ref{1.06a}) for the bulk metric defined by (\ref{1.01}), we first find the non-vanishing Christoffel symbols upto $\epsilon^3$ order. These are as listed as follows.

For the flat Rindler background $g_{a b}^{(0)}$ at $\mathcal{O}(\epsilon^0)$, the non-zero Christoffel symbols are:
\begin{align}
\Gamma^{r \, (0)}_{\tau r} = -\frac{1}{2} \, , \quad \Gamma^{r \, (0)}_{\tau \tau} = \frac{r}{2} \, , \quad \Gamma^{\tau \, (0)}_{\tau \tau} = \frac{1}{2} \, .
\label{A1.1}
\end{align}
For the perturbation $h_{a b}^{(3)}$ that kicks in at $\mathcal{O}(\epsilon^3)$, the sole non-zero Christoffel symbol is:
\begin{align}
\Gamma^{A \, (3)}_{\tau \tau} = \frac{r}{2} \, \delta^{A B} \bigg(\frac{a_1}{r_c} \partial_B P + a_2 \partial^2 v_B + \frac{a_3}{r_c} \partial_A v^2 \bigg) \, .
\label{A1.2}
\end{align}
Keeping in mind the scaling of the partial derivatives as given in (\ref{1.001}) and using (\ref{A1.1}) along with (\ref{A1.2}), we first evaluate the projected parallel transport equation with the free index assuming the time variable $\tau$, i.e., Eq. (\ref{1.05a}) order by order in the hydrodynamic expansion parameter $\epsilon$:
\begin{align}
&\mathcal{O}(\epsilon^0): \quad -r \Gamma^{\tau \, (0)}_{\tau \tau} + \Gamma^{r \, (0)}_{\tau \tau} = - \frac{r}{2} + \frac{r}{2} = 0 \,. \\
&\mathcal{O}(\epsilon^1): \quad -r \big{(}\Gamma^{\tau \, (1)}_{\tau \tau} + 2 \Gamma^{\tau \, (0)}_{\tau A} v^A\big{)} + \Gamma^{r \, (1)}_{\tau \tau} + 2 \Gamma^{r \, (0)}_{\tau A} v^A= 0 \,. \\
&\mathcal{O}(\epsilon^2): \quad -r \big{(}\Gamma^{\tau \, (2)}_{\tau \tau} + 2 \Gamma^{\tau \, (1)}_{\tau A} v^A + \Gamma^{\tau \, (0)}_{A B} v^A v^B\big{)} + \Gamma^{r \, (2)}_{\tau \tau} + 2 \Gamma^{r \, (1)}_{\tau A} v^A + \Gamma^{r \, (0)}_{A B} v^A v^B = 0 \,. \\
&\mathcal{O}(\epsilon^3): \quad -r \big{(}\Gamma^{\tau \, (3)}_{\tau \tau} + 2 \Gamma^{\tau \, (2)}_{\tau A} v^A + \Gamma^{\tau \, (1)}_{A B} v^A v^B\big{)} + \Gamma^{r \, (3)}_{\tau \tau} + 2 \Gamma^{r \, (2)}_{\tau A} v^A + \Gamma^{r \, (1)}_{A B} v^A v^B = 0 \,.
\label{A1.3}
\end{align}
Thus, we find that (\ref{1.05a}) vanishes identically upto $\mathcal{O}(\epsilon^3)$ and that (\ref{1.04}) is satisfied trivially for $c = \tau$.

Now, we evaluate the projected parallel transport equation with the free index assuming the transverse angular variables $x^A$, i.e., Eq. (\ref{1.06a}) in a similar order by order fashion in $\epsilon$:
\begin{align}
&\mathcal{O}(\epsilon^0): \quad \Gamma^{B \, (0)}_{\tau \tau} = 0 \,. \\
&\mathcal{O}(\epsilon^1): \quad \Gamma^{B \, (1)}_{\tau \tau} + 2 \Gamma^{B \, (0)}_{C \tau} v^C = 0 \,. \\
&\mathcal{O}(\epsilon^2): \quad \Gamma^{B \, (2)}_{\tau \tau} + 2 \Gamma^{B \, (1)}_{C \tau} v^C + \Gamma^{B \, (0)}_{C D} v^C v^D = 0 \,. \\
&\mathcal{O}(\epsilon^3): \quad \delta_{A B} \big{(} \partial_{\tau} v^B + v^C \partial_C v^B + \Gamma^{B \, (3)}_{\tau \tau} + 2 \Gamma^{B \, (2)}_{C \tau} v^C + \Gamma^{B \, (1)}_{C D} v^C v^D \big{)} \bigg{|}_{r = r_c} \,. \nonumber \\
&\quad \quad \quad \quad = \delta_{A B} \bigg{[} \partial_{\tau} v^B + v^C \partial_C v^B + \delta^{B D} \frac{r}{2} \big{(} \frac{a_1}{r_c} \partial_D P + a_2 \partial^2 v_D + \frac{a_3}{r_c} \partial_D v^2 \big{)} \bigg{]} \bigg{|}_{r = r_c} \nonumber \\
&\quad \quad \quad \quad = \partial_{\tau} v_A + v^C \partial_C v_A + \frac{a_1}{2} \partial_A P + \frac{a_2}{2} r_c \partial^2 v_A + \frac{a_3}{2} \partial_A v^2 \,. 
\label{A1.4}
\end{align}
So the satisfaction of Eq. (\ref{1.04}) identically upto all orders in $\epsilon$ requires that the above (\ref{A1.4}) must vanish at this order, which consequently leads to the fluid dynamical equations as discussed extensively in section {\ref{paraorder}}.  

\section{{\label{App3}}Evaluation of the expansion parameter $\theta$ (\ref{1.09a}) upto $\epsilon^3$ order}
As before, we proceed to explicitly check the right hand side of (\ref{1.09a}) in an order-wise manner in the hydrodynamic expansion parameter $\epsilon$:
\begin{align}
&\mathcal{O}(\epsilon^0): \quad h^{a b} (r \Gamma^{\tau \, (0)}_{a b} - \Gamma^{r \, (0)}_{a b}) = r h^{\tau \tau} \Gamma^{\tau \, (0)}_{\tau \tau} - h^{\tau \tau} \Gamma^{r \, (0)}_{\tau \tau} = r \bigg(-\frac{1}{r}\bigg) \frac{1}{2} - \bigg(-\frac{1}{r}\bigg) \frac{r}{2} = 0 \,. \\
&\mathcal{O}(\epsilon^1): \quad h^{a b} (r \Gamma^{\tau \, (1)}_{a b} - \Gamma^{r \, (1)}_{a b} - \Gamma^{A \, (0)}_{a b} v_A) = 0 \,. \\ \label{A2.1}
&\mathcal{O}(\epsilon^2): \quad h^{a b} (\partial_a v_b + r \Gamma^{\tau \, (2)}_{a b} - \Gamma^{r \, (2)}_{a b} - \Gamma^{A \, (1)}_{a b} v_A) = \delta^{A B} \partial_A v_B = \partial_A v^A \,. \\ \label {A2.2}
&\mathcal{O}(\epsilon^3): \quad h^{a b} (r \Gamma^{\tau \, (3)}_{a b} - \Gamma^{r \, (3)}_{a b} - \Gamma^{A \, (2)}_{a b} v_A) = 0 \,.
\end{align}
Thus, the vanishing of the expansion parameter $\theta$ as defined in (\ref{1.09a}) identically upto all orders in $\epsilon$ requires that the quantity in (\ref{A2.1}) must vanish at this order, which happens to be the incompressibility condition for the fluid flow, as discussed extensively in section {\ref{exp}}.


\end{document}